\documentclass[twocolumn,showpacs,aps,prl,superscriptaddress,amsmath,amssymb]{revtex4}

\usepackage{color}
\usepackage{graphicx}
\usepackage{dcolumn}
\usepackage{bm}
\usepackage{subfigure}
\usepackage{multirow}

\newcommand{\BABARPubYear}       {13}
\newcommand{\BABARPubNumber}     {011}
\newcommand{\SLACPubNumber} {15515}

\input{babarsym}

\long\def\inst#1{\par\nobreak\kern 4pt\nobreak
    {\it #1}\par\vskip 10pt plus 3pt minus 3pt}

\begin{document}

{\pagestyle{empty}

\preprint{\babar-PUB-\BABARPubYear/\BABARPubNumber}
\preprint{SLAC-PUB-\SLACPubNumber} 

\begin{flushleft}
\babar-PUB-\BABARPubYear/\BABARPubNumber \\
SLAC-PUB-\SLACPubNumber
\end{flushleft}

\title{\boldmath{Measurement of the Mass of the $D^{0}$ Meson}}\vspace{0.3in}
%
\author{J.~P.~Lees}
\author{V.~Poireau}
\author{V.~Tisserand}
\affiliation{Laboratoire d'Annecy-le-Vieux de Physique des Particules (LAPP), Universit\'e de Savoie, CNRS/IN2P3,  F-74941 Annecy-Le-Vieux, France}
\author{E.~Grauges}
\affiliation{Universitat de Barcelona, Facultat de Fisica, Departament ECM, E-08028 Barcelona, Spain }
\author{A.~Palano$^{ab}$ }
\affiliation{INFN Sezione di Bari$^{a}$; Dipartimento di Fisica, Universit\`a di Bari$^{b}$, I-70126 Bari, Italy }
\author{G.~Eigen}
\author{B.~Stugu}
\affiliation{University of Bergen, Institute of Physics, N-5007 Bergen, Norway }
\author{D.~N.~Brown}
\author{L.~T.~Kerth}
\author{Yu.~G.~Kolomensky}
\author{M.~J.~Lee}
\author{G.~Lynch}
\affiliation{Lawrence Berkeley National Laboratory and University of California, Berkeley, California 94720, USA }
% ...
\collaboration{The \babar\ Collaboration}
\noaffiliation

\begin{abstract}
We report a measurement of the $D^0$ meson mass using the decay chain $D^{*}\left(2010\right)^{+} \rightarrow D^0 \pi^+$ with $D^0 \rightarrow K^- K^- K^+ \pi^+$. The data were recorded with the \babar\ detector at center-of-mass energies at and near the $\Upsilon(4S)$ resonance, and correspond to an integrated luminosity of approximately $477 \invfb$. We obtain $m(D^0) = (1864.841 \pm 0.048 \pm 0.063) \mev$, where the quoted errors are statistical and systematic, respectively. The uncertainty of this measurement is half that of the best previous measurement.
\end{abstract}


\maketitle                                                                             
}

\setcounter{footnote}{0}

\section{Introduction}\label{sec:intro}
The $D^0$ is one of the ground states of the charm mesons, and its mass sets the mass scale for the heavier excited states. As such, the $D^0$ mass is directly relevant to several measurements. For example, the reported $D^{*}\left(2010\right)^{+}$ mass is the sum of the nominal $D^0$ mass and the measured difference, $\Delta m$, between the masses of the $D^{*+}$ and $D^0$ mesons. Mixing parameters in the $D^0-\Dbar^0$ system can be extracted from a time-dependent amplitude analysis of the mass-constrained Dalitz plot, as in the analyses of the $K_{S}^0\pi^+\pi^-$ and $K_{S}^0 K^+ K^-$ final states \cite{PhysRevLett.105.081803,PhysRevD.72.012001,PhysRevLett.99.131803}. A new value of the $D^0$ mass would shift an event's position in the Dalitz plot if the final state is constrained to the $D^0$ mass value. A precise $D^0$ mass measurement can also serve as a reference point in magnetic field calibration studies. The $D^0$ mass is also important for the determination of the relationship of the $D^0 \Dbar^{*0}$ threshold to the mass of $X(3872)$. In this regard, a recent analysis by LHCb~\cite{PhysRevLett.110.222001} reported $J^{PC} = 1^{++}$ for $X(3872)$, which favors an exotic model such as one in which the $X(3872)$ is a $D^0 \Dbar^{*0}$ molecule \cite{Tornqvist:2004qy}.

We measure the $D^0$ mass using the decay chain $D^{*}\left(2010\right)^{+} \rightarrow D^0 \pi^+$, $D^0 \rightarrow K^- K^- K^+ \pi^+$. The use of charge conjugate reactions is implied here and throughout this paper. We chose this mode specifically for the relatively low Q-value, $m(K^- K^- K^+ \pi^+) - 3m(K^\pm) - m(\pi^+) \approx 250 \mev$, which produces small backgrounds, yields good resolution, and minimizes systematic uncertainties.  The previous most precise measurements were made by the CLEO collaboration~\cite{cleo}, which reported $m(D^0) = (1864.847 \pm 0.150 \pm 0.095) \mev$, and by the LHCb collaboration \cite{Aaij:2013uaa}, which reported $m(D^0) = (1864.75 \pm 0.15 \pm 0.11) \mev$, where the errors are statistical and systematic, respectively. The integrated luminosity for this analysis is about 60 times that of the CLEO analysis. After selection criteria are chosen to minimize systematic uncertainties, our sample is about 15 times larger than the CLEO sample \cite{cleo}. The LHCb analysis \cite{Aaij:2013uaa} uses the same decay channel we use, but their signal has about half the number of events, and our signal-to-background ratio is much higher.

\section{Data Sample and Detector}\label{sec:detector}
This analysis is based on a data sample corresponding to an integrated luminosity of approximately 477 fb$^{-1}$ \cite{Lees:2013rw} recorded at, and $40 \mev$ below, the $\Upsilon(4S)$ resonance by the \babar\ detector at the PEP-II asymmetric energy \epem\ collider at the SLAC National Accelerator Laboratory. The \babar\ detector is described in detail elsewhere~\cite{original_nim, babarnim}. Here, we summarize the most relevant components. The momenta of charged particles are measured with a combination of a cylindrical drift chamber (DCH) and a 5-layer silicon vertex tracker (SVT), both operating within the $1.5$ T magnetic field of a superconducting solenoid. Information from a ring-imaging Cherenkov detector is combined with specific ionization $(dE/dx)$ measurements from the SVT and DCH to identify charged-kaon and pion candidates. Electrons are identified, and photons measured, with a CsI(Tl) electromagnetic calorimeter. The return yoke of the superconducting coil is instrumented with tracking chambers for the identification of muons.

\section{Event Selection}\label{sec:selection}

We expect our measurement to be dominated by systematic uncertainties, so we focus on choosing a balanced set of selection criteria that produces a very clean signal, to control systematic uncertainties, while still preserving a relatively large number of signal events. To avoid potential bias, the final selection criteria are chosen based on extensive studies of Monte Carlo (MC) simulation events and on 5\% of the real data; the latter are not used in the mass measurement. With the final selection criteria determined from these preliminary studies, systematic uncertainties are determined from a blind analysis of the remaining 95\% of the data, in which the measured mass of each event is shifted by an unknown, common offset. After completion of all the systematic uncertainty studies, we removed the offset and performed a fit to the measured mass spectrum.

We reconstruct the decay chain $D^{*}\left(2010\right)^{+} \rightarrow D^0 \pi^+$, $D^0 \rightarrow K^- K^- K^+ \pi^+$ using a kinematic fit requiring that the $D^0$ daughters emerge from a common vertex and that the $D^0$ and the $D^{*}\left(2010\right)^{+}$ ($D^{*+}$)  daughter pion momenta point back to the primary vertex, with the additional constraint that the $D^{*+}$ candidates originate from the luminous region~\cite{Asner:2000mb}. The $ \chi^2 / \nu $ ($ \nu $ denotes the number of degrees of freedom) reported by the fit was required to be less than 20/9, corresponding to a fit probability greater than 1.8\%. For events with multiple candidates, we choose the candidate with the largest confidence level.

To produce a good balance between expected statistical precision and systematic uncertainty we apply a variety of selection criteria. We consider the momentum of the $D^{*+}$ in the \epem\ center-of-mass frame, $p^*(D^{*+})$, and $\Delta m$. In addition, we require kaon and pion tracks to pass particle identification (PID) selections. We look at the statistical significance and purity of data sets with varying $p^*(D^{*+})$, $\Delta m$, and PID cuts. We choose a reasonable set of selection criteria, $p^*(D^{*+}) > 2.5 \gev$, $\left|\Delta m - \Delta m_{\rm PDG}\right| < 1.5 \mev$, where the PDG subscript indicates the value listed by the PDG \cite{pdg2012}, before the remaining 95\% of the data sample is studied, thus avoiding potential analyst bias.

In approximately 1\% of events, we mis-reconstruct the decay and swap the pion from the $D^{*+}$ decay (referred to as the slow pion, $\pi_{s}$) with the pion from the $D^{0}$ decay. This exchange could potentially shift the measured mass value, as these events tend to concentrate at lower values of reconstructed $D^0$ mass. To eliminate these events, we define the variables $m'(D^0) = m(K^- K^- K^+ \pi_{s}^+)$ and $\Delta m' = m(K^- K^- K^+ \pi^+ \pi^+_{s}) - m(K^- K^- K^+ \pi^+_{s})$. In this ($m'$, $\Delta m'$) system, correctly reconstructed events are shifted away from the ($m'$, $\Delta m'$) signal region, and events with the exchanged pion mis-reconstruction are shifted into the ($m'$, $\Delta m'$) signal region. The requirement $\Delta m' > 0.15 \gev$ eliminates 80\% of the exchanged events that survive as $D^0$ candidates with essentially no loss of correctly reconstructed signal. According to MC simulation, the surviving 20\% of mis-reconstructed events have daughter-pion and slow-pion momenta so similar that the wrong mass values have a negligible  effect on our $ D^0 $ mass determination.

In a prior study of the line shape and relativistic Breit-Wigner pole position in $D^{*+} \rightarrow D^0 \pi^{+}_{s}$ decays \cite{zachspaper}, we observed strong mass and mass-difference dependence on track momentum. We also observed that the reconstructed $K^0_S$ mass value was systematically low and varied with momentum. This effect was not replicated in MC. 

We determined that increasing the magnitude of the laboratory momentum by a factor of 1.0003 and increasing the energy loss reported by the Kalman fit by 1.8\% and 5.9\% for the beam pipe and SVT, respectively, removes the momentum dependence and offset of the reconstructed $K_S^0$ mass. The increase in the magnitude of momentum corresponds to increasing the magnetic field by 0.45 mT. The corrections also eliminated the momentum dependence of the mass difference, which was the subject of the $D^{*+} \rightarrow D^0 \pi^{+}_{s}$ study. The procedure for determining these parameters and those used for systematic uncertainty corrections is detailed in Ref.~\cite{zachspaper}. We apply the same magnetic-field and material-model-corrections in this analysis.

In studying the $K_S^0$ signal, we observed that even after the corrections described above, the $K_S^0$ mass dropped dramatically when either of its daughter tracks had $\cos \theta > 0.89$. We therefore reject events for which any of the daughter tracks of the $D^0$ has $\cos \theta > 0.89$. This criterion reduces the final data sample by approximately 10\%. We additionally require that the momentum of the slow pion is at least $150 \mev$ in the laboratory frame in order that the MC laboratory momentum distribution accurately replicates that observed for data.

\section{Fit to Data}\label{sec:fit}
\begin{figure}[!h]
\begin{center}
\includegraphics[scale=0.4]{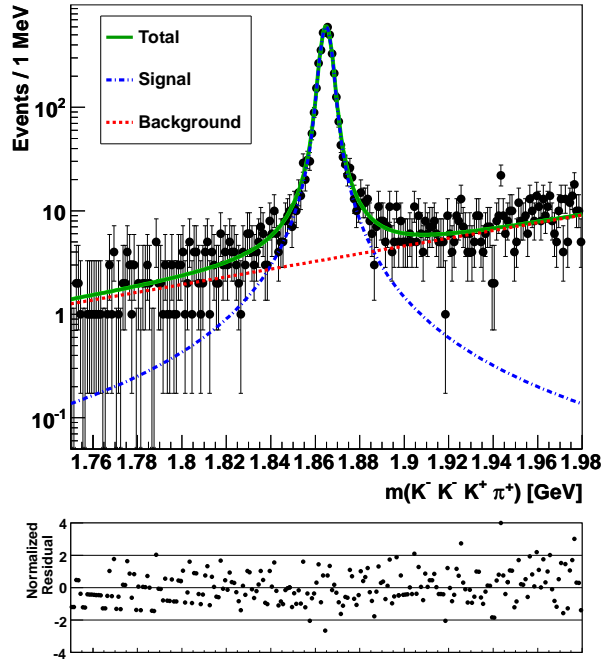}
\caption{(color online) Results of the extended, unbinned maximum likelihood fit to the reconstructed $D^0$ mass distribution. Normalized residuals are defined as $(N_o-N_p)/\sqrt{N_p}$, where $N_o$ and $N_p$ are observed and predicted numbers of events, respectively.}
\label{fig:fit}
\end{center}
\end{figure}

To determine the $D^0$ mass, we perform an extended, unbinned maximum likelihood fit over the candidate mass range $1.75 - 1.98 \gev$, using a Voigtian distribution to describe the signal. This function provides better agreement between the model and the $m(K^- K^- K^+ \pi^+)$ signal distribution than a multi-Gaussian model. The background is described by an exponential function. The Voigtian distribution is the convolution of a Cauchy and a Gaussian distribution, defined as

\begin{equation}
V(m) = \int_{-\infty}^{\infty} \mathrm{d} \mu \exp\left[\frac{-(\mu - m_D)^2}{2 \sigma^2}\right] \frac{1}{(m - \mu)^2+ \frac{\gamma^2}{4}},
\end{equation}

\noindent where $m$ is the $K^-K^-K^+\pi^+$ invariant mass, and the three parameters are $m_D$, the $D^0$ mass, and $\gamma$ and $\sigma$, two ad-hoc resolution parameters. The fit to the data is shown in Fig.~\ref{fig:fit}, and the results of this fit are summarized in Table~\ref{tab:fitstats}. At the peak, the signal-to-background ratio is approximately 175:1.

\begin{table}[!h]
\begin{center}
\caption{The results of the fits to data for the $K^- K^- K^+ \pi^+$ channel (statistical uncertainties only).}
\begin{tabular}{c c}
\hline \hline \\[-1.9ex]
Fit Parameters & Values \\[-1.9ex] \\ \hline \\[-1.7ex]
Number of signal events & 4345 $\pm$ 70\\
$m(D^0)$ (\mev) & 1864.841 $\pm$ 0.048\\
$\gamma$ (\mev) & 2.596 $\pm$ 0.152 \\
$\sigma$ (\mev) & 1.762 $\pm$ 0.086\\\\[-1.7ex]
\hline \hline
\end{tabular}
\label{tab:fitstats}
\end{center}
\end{table}

\section{Systematic Uncertainties} \label{sec:syserr}

Systematic uncertainties arise from a variety of sources. The dominant systematic uncertainty is the effect of the uncertainty in the charged-kaon mass on the $K^-K^-K^+\pi^+$ invariant mass. The corrections to the momentum scale and $dE/dx$ energy loss in detector material are also varied to account for the uncertainty in the $K_{S}^{0}$ mass \cite{zachspaper, pdg2012}. Additionally, we vary the fit model to estimate the systematic uncertainty associated with our choice of signal and background shape, and we study the systematic uncertainty associated with inner bremsstrahlung. We then divide the data into disjoint subsets corresponding to intervals of azimuthal angle, $\phi$, the laboratory momentum of the $D^0$, $p_{\text{lab}}$, and $\Delta m$. We assign systematic uncertainties using a method similar to that used in obtaining the PDG scale factor \cite{pdg2012}, as described below. The systematic uncertainties are summarized in Table~\ref{tab:syssummary}.

\begin{table}
\begin{center}
\caption{Systematic uncertainties for the $D^0$ mass measurement.}
\begin{tabular}{ c  c }
\hline \hline \\[-1.9ex]
Source & $\sigma_{\text{sys}}$ (\mev) \\[-1.9ex] \\ \hline \\[-1.7ex]
$K^{\pm}$ mass uncertainty & 0.046 \\
Magnetic field and material model & 0.031 \\
Signal shape & 0.009 \\
Background model & 0.005\\
Bremsstrahlung  & $< 0.001$\\
Disjoint $\phi$ interval variation &  0.000\\
Disjoint $p_{\text{lab}}$ interval variation &  0.000\\ 
Disjoint $\Delta m$ interval variation & 0.028\\ \\[-1.7ex] \hline \\[-1.9ex]
Sum in quadrature & 0.063 \\[-1.9ex] \\ \hline \hline
\end{tabular}
\label{tab:syssummary}
\end{center}
\end{table}

To estimate the effect of the charged-kaon mass uncertainty on the reconstructed $D^0$ mass, we vary the charged-kaon mass value by $\pm 1 \sigma_{\text{PDG}} = \pm 16 \kev$ \cite{pdg2012}, redetermine $m(K^- K^- K^+ \pi^+)$ of each event, and refit each new $m(K^- K^- K^+ \pi^+)$ distribution. We take the average of the magnitude of the differences from the nominal fit results as the systematic uncertainty. We find an average variation of $46 \kev$, which proves to be the largest source of systematic uncertainty. The charged pion mass is known to a much higher precision, so its mass uncertainty has a negligible effect on the reconstructed $D^0$ mass value.

We estimate the uncertainty associated with the choice of our correction parameters for the magnetic field and detector material model by examining the variation between the fit results using the nominal parameter values and those obtained by tuning to the mass values $m_{\text{PDG}}\left(K_{S}^{0}\right)\pm 1\sigma_{\text{PDG}}$ \cite{zachspaper, pdg2012}. We compare the fitted mass value from the nominal fit to the mass values extracted using the $\pm 1 \sigma_{\text{PDG}}$ correction parameters and take the largest difference between those fits and the nominal fit, $31 \kev$, as a measure of the systematic uncertainty.

We alternatively fit the real data with a double Gaussian signal shape and see a shift of $9 \kev$ in the central mass value. We also change the background model used in the fit procedure from the nominal exponential distribution to a second degree polynomial. Although it provides a relatively poor description, using a quadratic background changes the central value by only $5 \kev$. We report the magnitudes of these shifts as systematic uncertainties in Table~\ref{tab:syssummary}.

We study the systematic uncertainty associated with inner bremsstrahlung using PHOTOS \cite{Barberio:1993qi}. About 3.5\% of the generated $ D^0 \to K^- K^- K^+ \pi^+ \, n \, \gamma $ decays had at least one photon, and in these events the $K^- K^- K^+ \pi^+$ invariant mass averaged about $15 \kev$ below the nominal $D^0$ mass, leading to a shift in the full sample average of less than $1 \kev$; we therefore neglect this as a source of systematic uncertainty.

\begin{figure}[h]
\begin{center}
\subfigure{\label{fig:phi_dep}\includegraphics[scale=0.4]{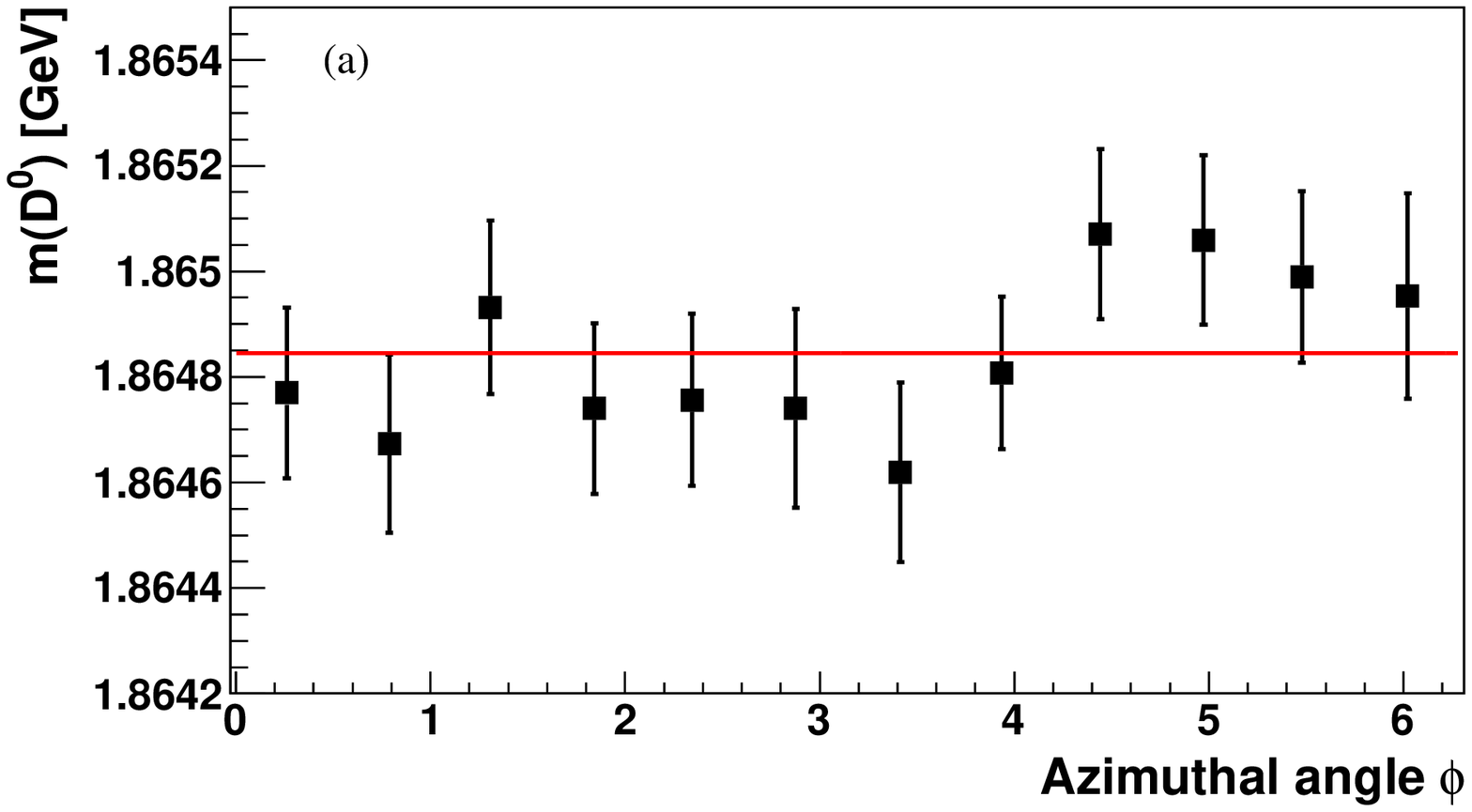}}
\subfigure{\label{fig:plab_dep}\includegraphics[scale=0.4]{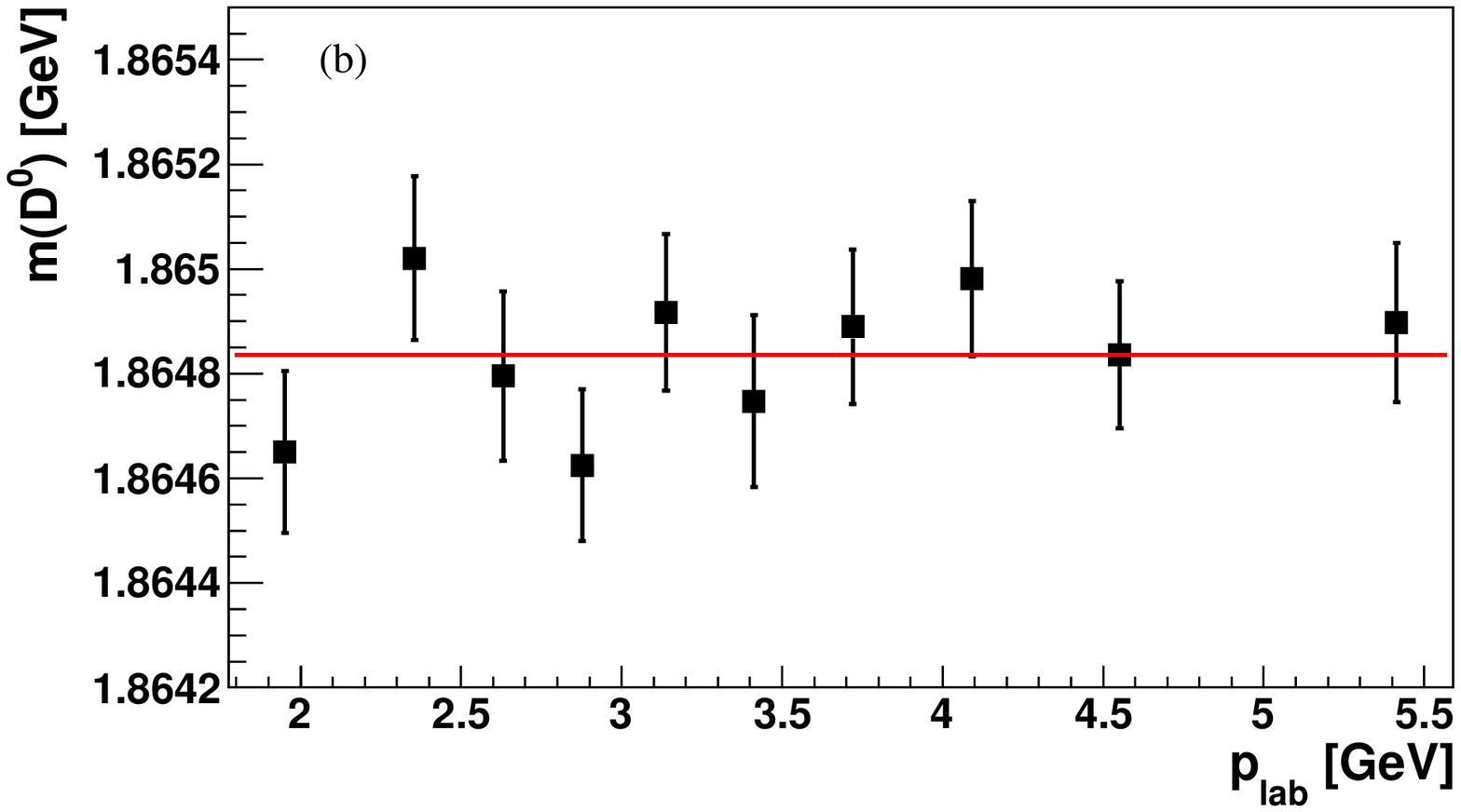}}
\subfigure{\label{fig:dm_dep}\includegraphics[scale=0.4]{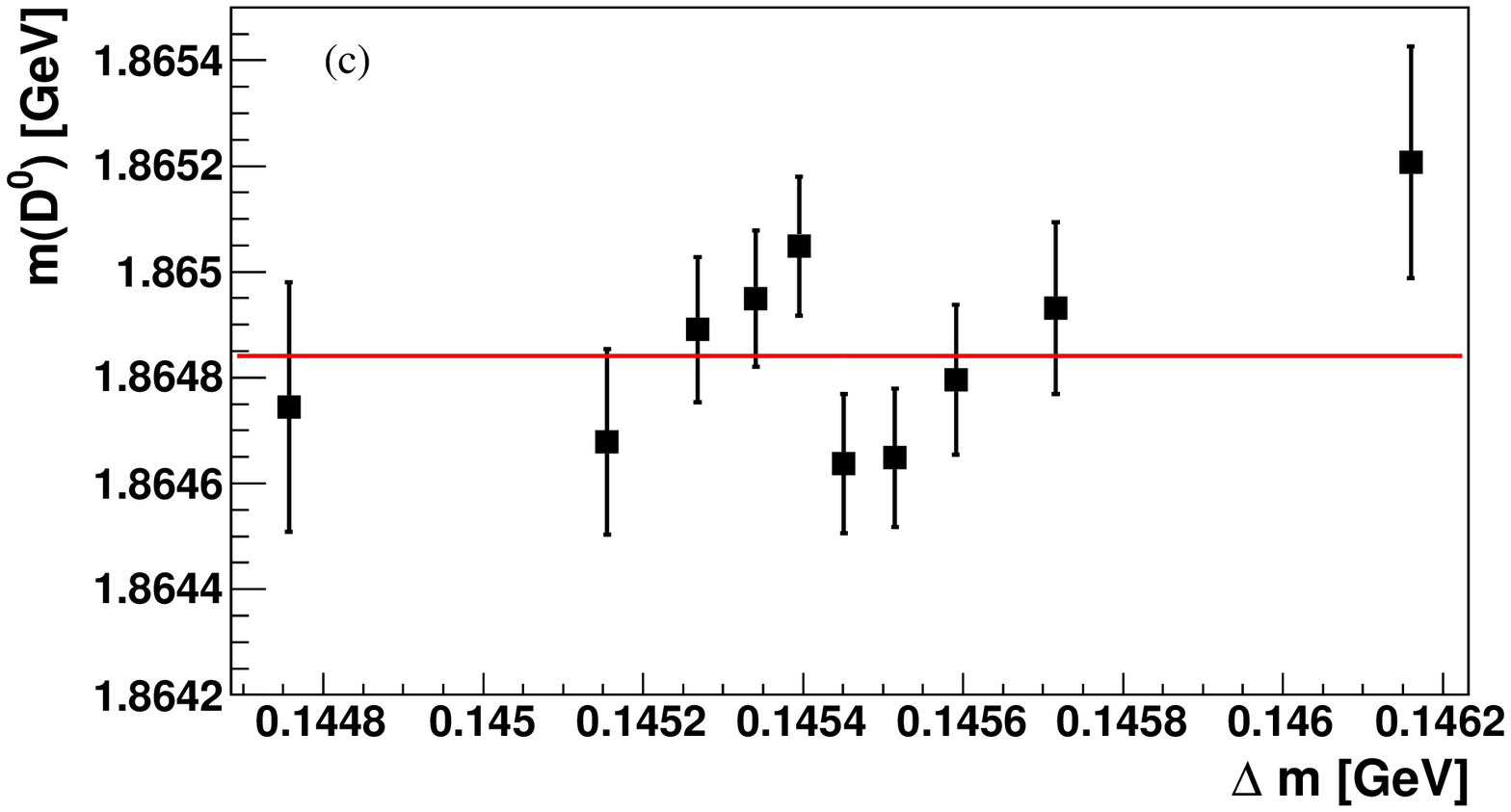}}
\caption{
The value of $m(D^0)$ obtained from fits to data divided into disjoint subsets in \subref{fig:phi_dep} $\phi$, \subref{fig:plab_dep} $p_{\text{lab}}(D^0)$, and \subref{fig:dm_dep} $\Delta m$. Each point represents an individual fit, and the horizontal line indicates the nominal fit result; the points are plotted at the mean value in the range, and the uncertainty in the mean values is smaller than the marker size.}
\label{fig:syst}
\end{center}
\end{figure}

We study the $D^0$ mass dependence on $p_{\rm lab}(D^0)$, on $\Delta m$, and on $\phi$, based on dependences observed in previous \babar\ analyses. We divide the data into 10 subsets for $p_{\text{lab}}$ and $\Delta m$, and 12 subsets for $\phi$ (to reflect the 6-fold symmetry of the detector); the intervals in the $ p_{\rm lab} $ and $ \Delta m $ plots have variable widths selected to include roughly equal numbers of signal events. If the fit results from the disjoint subsets are compatible with a constant value, in the sense that $\chi^2/\nu < 1$, we assign no systematic uncertainty. However, if we find $\chi^2/\nu > 1$ we ascribe an uncertainty using a variation on the scale factor method used by the PDG (see the discussion of unconstrained averaging~\cite{pdg2012}). In our version of this procedure, we determine the systematic uncertainty for unknown detector issues to be the value that, when summed in quadrature with the statistical error, produces $ \chi^2 / \nu = 1 $, so that

\begin{align}
\label{eq:syserror}
\sigma_{\text{sys}}  &= \sigma_{\text{stat}} \sqrt{S^2 - 1} ,
\end{align}

\noindent where $S^2 = \chi^2/\nu$. The $\chi^2$ statistic gives a measure of fluctuations, including those expected from statistics, and those from systematic effects. Once we remove statistical fluctuations, we associate what remains with a possible systematic uncertainty. Fig.~\ref{fig:syst} does not show any obvious patterns of variations, but using this procedure we assign a $28 \kev$ systematic uncertainty due to the variations seen in the $\Delta m$ subsets. We find that the variation of the $D^0$ mass with azimuthal angle and with laboratory momentum is consistent with statistical fluctuations, and thus assign no associated systematic error.

\section{Conclusion}\label{sec:conclusion}

We have measured the $D^0$ mass with more than twice the precision of the previous best measurement~\cite{cleo} by analyzing a high-purity sample of $D^{*+}$ continuum events produced in \epem\ collisions near $10.6 \gev$. The corresponding integrated luminosity is approximately $477 \invfb$. Reconstructing the decay $D^0 \rightarrow K^- K^- K^+ \pi^+$, we measure $m(D^0) = (1864.841 \pm 0.048 \pm 0.063) \mev$. The dominant source of systematic uncertainty is the uncertainty in the charged-kaon mass. This measurement will significantly improve the precision of the current world average, $(1864.86 \pm 0.13) \mev$ \cite{pdg2012}.

The largest source of uncertainty in the present measurement comes from the charged-kaon mass uncertainty, reported to be $\pm 16 \kev$ by the PDG \cite{pdg2012}. The two most precise measurements of the charged-kaon mass \cite{Denisov:1991pu, Gall:1988ei} are both more than one standard deviation from the PDG central value, and they differ by $60 \kev$ with reported uncertainties of only $7 \kev$ and $11 \kev$. Therefore, we also give the dependence of our final result on $m(K^\pm)$, $m(D^0) = (1864.841 \pm 0.048 \pm 0.043 + 3[m(K^\pm) - 493.677]) \mev$, where the uncertainties are statistical and instrumental, respectively. Using this value, the $D^0$ mass can readily be obtained from an improved kaon mass value.

We may use our result to estimate the binding energy of the $X(3872)$, interpreting it as a $D^0 \Dbar^{*0}$ molecule. Combining it with the current PDG average~\cite{pdg2012} for the $D^{*}(2007)^{0}-D^{0}$ mass difference, $(142.12 \pm 0.07) \mev$, and the PDG average mass of the $X(3872)$, $(3871.68 \pm 0.17) \mev$, yields 

\begin{align*}
E_b &= m(D^0)+m(D^{*0})-m(X(3872)) \\
& = 2 m(D^0) + \left[m(D^{*0})-m(D^{0})\right] - m(X(3872))\\
& = (0.12 \pm 0.24) \mev.
\end{align*}

\section{Acknowledgements}
We are grateful for the 
extraordinary contributions of our \pep2\ colleagues in
achieving the excellent luminosity and machine conditions
that have made this work possible.
The success of this project also relies critically on the 
expertise and dedication of the computing organizations that 
support \babar.
The collaborating institutions wish to thank 
SLAC for its support and the kind hospitality extended to them. 
This work is supported by the
US Department of Energy
and National Science Foundation, the
Natural Sciences and Engineering Research Council (Canada),
the Commissariat \`a l'Energie Atomique and
Institut National de Physique Nucl\'eaire et de Physique des Particules
(France), the
Bundesministerium f\"ur Bildung und Forschung and
Deutsche Forschungsgemeinschaft
(Germany), the
Istituto Nazionale di Fisica Nucleare (Italy),
the Foundation for Fundamental Research on Matter (The Netherlands),
the Research Council of Norway, the
Ministry of Education and Science of the Russian Federation, 
Ministerio de Ciencia e Innovaci\'on (Spain), and the
Science and Technology Facilities Council (United Kingdom).
Individuals have received support from 
the Marie-Curie IEF program (European Union) and the A. P. Sloan Foundation (USA). 
The University of Cincinnati is gratefully acknowledged for its support of 
this research through a WISE (Women in Science and Engineering) fellowship to C. Pappenheimer.



\begin{thebibliography}{99}
\bibitem{PhysRevLett.105.081803} P. del Amo Sanchez {\em et al.} (\babar\ Collaboration), Phys. Rev. Lett. {\bf 105}, 081803 (2010).
\bibitem{PhysRevD.72.012001} D. M. Asner {\em et al.} (CLEO Collaboration), Phys. Rev. D {\bf 72}, 012001 (2005).
\bibitem{PhysRevLett.99.131803} L. M. Zhang {\em et al.} (Belle Collaboration), Phys. Rev. Lett. {\bf 99}, 131803 (2007).
\bibitem{PhysRevLett.110.222001} R. Aaij {\em et al.} (LHCb Collaboration), Phys. Rev. Lett. {\bf 110}, 222001 (2013).
\bibitem{Tornqvist:2004qy} N. A. Tornqvist, Phys. Lett. B {\bf 590}, 209 (2004).
\bibitem{cleo} C. Cawlfield {\em et al.} (CLEO Collaboration), Phys. Rev. Lett. {\bf 98} 092002 (2007).
\bibitem{Aaij:2013uaa} R. Aaij {\em et al.} (LHCb Collaboration), JHEP {\bf 1306}, 065 (2013).
\bibitem{Lees:2013rw} J.~P.~Lees {\it{et al.}} (\babar\ Collaboration), Nucl. Instr. Methods Phys. Res., Sect. A {\bf{726}}, 203 (2013).
\bibitem{original_nim} B. Aubert {\it{et al.}} (\babar\ Collaboration), Nucl. Instr. Methods Phys. Res., Sect. A {\bf{479}} 1 (2002).
\bibitem{babarnim} B. Aubert {\it{et al.}} (\babar\ Collaboration), in press in Nucl. Instr. Methods Phys. Res., Sect. A arXiv:1305.3560 (2013).
\bibitem{Asner:2000mb} D. Asner, Ph.D. thesis, Univesity of California at Santa Barbara 2000.
\bibitem{pdg2012} J. Beringer {\em et al.} (Particle Data Group), Phys. Rev. D {\bf 86} 010001 (2012).
\bibitem{zachspaper} J. P. Lees {\em et al.} (\babar\ Collaboration), arXiv:1304.5009.
\bibitem{Barberio:1993qi} E. Barberio and Z. Was, Comput. Phys. Commun. {\bf 79}, 291 (1994).
\bibitem{Denisov:1991pu} A. Denisov, A. Zhelamkov, Y. Ivanov, L. Lapina, P. Levchenko, {\em et al.}, JETP Lett. {\bf 54} 558 (1991). 
\bibitem{Gall:1988ei} K. Gall, E. Austin, J. Miller, F. O'Brien, B. Roberts {\em et al.}, Phys. Rev. Lett. {\bf 60}, 186 (1988).
\end{thebibliography}
\end{document}